\documentstyle[12pt]{article}
\begin{document}

\vspace*{1cm}
\begin{center}
{\large\bf Hartree-Fock ground state of the composite fermion metal}
\end{center}

\vspace{1cm}

\noindent Piotr Sitko and Lucjan Jacak, Institute of Physics,
Technical University of Wroc\l{}aw,\\ Wyb. Wyspia\'{n}skiego
 27, 50-370 Wroc\l{}aw, Poland.

 \vspace{1cm}
\begin{center}

Abstract

\end{center}
Within the Hartree-Fock approximation the ground state
of the composite fermion metal is found.
We observe that the single-particle energy spectrum is dominated by the
logarithmic interaction exchange term which leads to
an infinite jump of the single-particle energy
at the Fermi momentum. It is shown that the Hartree-Fock result brings
no corrections to the RPA Fermi velocity.

\vspace{2cm}
\newpage
\noindent{\bf  1. Introduction}
\vspace{0.5cm}

Recently there has appeared a great interest, both theoretical and
experimental, in studies of the compressible $\nu=1/2$ quantum Hall
state. Halperin, Lee and Read (HLR) \cite{HLR} constructed the theory of
quantum Hall states at fractions with even denominators.
The idea comes from the Jain theory of composite fermions originally	stated
in the context of the fractional quantum Hall effect (FQHE)
\cite{Jain}. In close
analogy to the Laughlin wave function Jain proposed simpler and more
elegant way of understanding the FQHE phenomenon which consists in attaching
an even number ($2f$) of flux quanta to each electron.
If one takes an average of fluxes (mean field approximation)
will find a partial cancelation of the external magnetic field leading
to new incompressible states which results in the FQHE.
As it was noticed in Ref. \cite{HLR}
such a construction at $\nu=1/(2f)$
leads to a metal-like state (exact cancelation of fields, i.e. the zero
effective field) which
is to some extent	verified experimentaly \cite{experiment}.

The full description of the composite-fermions picture is given in
terms of the Chern-Simons theory \cite{Lopez}.
Despite the mean field there are
 interaction terms including also a three-body term
which dominating part is a logarithmic interaction
\cite{Chen}.
HLR studied the linear response of the system within the random-phase
approximation.
They found that the Fermi velocity vanishes so that the  effective mass is
divergent which is in agreement with recent experiments \cite{rex}.
However, it is still uncertain whether the self-energy graphs
not included in the
RPA will normalize the effective mass \cite{renorm,kim}.
In this paper we consider
all first order contributions to the single-particle energy within the
Hartree-Fock (H-F) approximation.

\newpage
\noindent{\bf  2. Hartree-Fock ground state}
\vspace{0.5cm}

The starting point of our consideration is the Hamiltonian:
\begin{equation}
H=\frac{1}{2m}\sum_{i=1}^{N}({\bf P}_{i}+\frac{e}{c}{\bf
A}_{i}+\frac{e}{c}{\bf
A}_{i}^{ex})^{2}
\end{equation}
where	the vector potential from the attached fluxes (Chern-Simons field)
is given by
\begin{equation}
{\bf A}_{i}=\sum_{j}{\bf A}_{ij}=\frac{-2f \hbar c}{e}	\hat{\bf z}\times
\sum_{j\neq i}
\frac{({\bf r}_{i}-{\bf r}_{j})}{|{\bf r}_{i}-{\bf r}_{j}|^{2}}\;\; ,
\end{equation}
$\hat{\bf z}$ - a perpendicular to the plane unit vector.
The Hamiltonian (1) for  integer values of $f$ describes
 composite fermions  in the external magnetic field and for
other values of $f$ -- anyons with  statistics
parameters
$\Theta = (1+2f)\pi$ (an exchange of two anyons produces a phase
factor of $e^{i\Theta}$).

The mean field $B^{s}$ is
given by the average vector potential:
\begin{equation}
\bar{\bf A}_{i}=\sum_{\bf p}\int |\phi_{\bf p}({\bf r}_{j})|^{2}{\bf
A}_{ij} d^{2}{\bf r}_{j}.
\end{equation}
When the effective field is zero, i.e. $B^{s}+B^{ex}=0$,
we assume that the  ground state of the system is given by
the plane wave: $\phi_{\bf p}({\bf
r}_{j})=\frac{1}{\sqrt{S}}\exp({\frac{i}{\hbar}{\bf p}{\bf r}_{j}})$.
Let us recall the expressions for the H-F ground-state energy
\cite{HFA,halffil}:
\begin{equation}
<H_{a}>=\frac{1}{2m}\sum_{\bf p}\int d1 \phi_{\bf p}^{+}(1)({\bf
P}_{1}+\bar{\bf A}_{1}+{\bf A}_{1}^{ex})^{2}\phi_{\bf p}(1)\;\; ,
\end{equation}
\begin{equation}
<H_{b}>=-\frac{e}{mc}\sum_{{\bf p},{\bf q}} \int d1d2\phi_{\bf
p}^{+}(1)\phi_{\bf q}^{+}(2){\bf A}_{12}({\bf P}_{1}+\bar{\bf A}_{1}+{\bf
A}_{1}^{ex})\phi_{\bf p} (2)
\phi_{\bf q} (1)\; ,
\end{equation}
\begin{equation}
<H_{c}>=\frac{e^{2}}{2c^{2}m}\sum_{{\bf p},{\bf q}}  \int d1d2\phi_{\bf
p}^{+}(1)\phi_{\bf q}^{+}(2)|{\bf A}_{12}|^{2}\phi_{\bf p}(1)
\phi_{\bf q}(2)\; ,
\end{equation}
\begin{equation}
<H_{d}>=-\frac{e^{2}}{2c^{2}m}\sum_{{\bf p},{\bf q}}\int d1d2\phi_{\bf
p}^{+}(1)\phi_{\bf q}^{+}(2)|{\bf A}_{12}|^{2}\phi_{\bf p}(2)
\phi_{\bf q}(1)\; ,
\end{equation}
\begin{equation}
<H_{e}>=\frac{e^{2}}{c^{2}m}\sum_{{\bf p},{\bf q},{\bf k}}\int
d1d2d3
\phi_{\bf p}^{+}(1)\phi_{\bf q}^{+}(2)\phi_{\bf k}^{+}(3)
{\bf A}_{12}{\bf A}_{13}
\phi_{\bf p}(3)\phi_{\bf q}(1)\phi_{\bf k}(2)\; ,
\end{equation}
\begin{equation}
<H_{f}>=-\frac{e^{2}}{2c^{2}m}\sum_{{\bf p},{\bf q},{\bf k}}\int
d1d2d3
|\phi_{\bf p}(1)|^{2}\phi_{\bf q}^{+}(2)\phi_{\bf k}^{+}(3)
{\bf A}_{12}{\bf A}_{13}
\phi_{\bf q}(3)\phi_{\bf k}(2)
\end{equation}
where $d1=d^{2}{\bf r}_{1}$, sums extend over the Fermi sea and
omit  elements of two equal momentum variables.
It was shown in Ref. \cite{halffil} that the Hartree-Fock ground-state
energy $<H>=\sum_{i=a}^{f}<H_{i}>$
of the composite fermion metal is finite and equals:
\begin{equation}
\label{fq}
<H>=(1+3f^{2})\frac{N}{2}E_{F}
\end{equation}
which confirms that the ground state of the system is the Fermi
sea.
Let us note that the result	(\ref{fq}) agrees with the H-F
ground-state energy of composite fermions in the FQHE at filling
fractions $\nu=\frac{n}{2fn+1}$	\cite{fqhe}
in the limit $n\rightarrow\infty$ ($n$
is a number of occupied Landau levels in the effective field).

The single-particle Hamiltonian is found via the first variation in
$\phi_{\bf p}^{+}$ of  Eqs.(4-9) \cite{HFA}. From the variation of
$<H_{a}>$ we find:
\begin{equation}
H_{HF}^{1}\phi_{\bf p}(1)=\frac{1}{2m}\sum_{\bf q}\int d2 \phi_{\bf
q}^{+}(2)
({\bf P}_{1}+\bar{\bf A}_{1}+{\bf A}_{1}^{ex})^{2}
\phi_{\bf q}(2)\phi_{\bf p}(1)\;\; ,
\end{equation}
\begin{equation}
H_{HF}^{2}\phi_{\bf p}(1)=\frac{e}{mc}\sum_{\bf q}\int d2 \phi_{\bf
q}^{+}(2)
{\bf A}_{21}({\bf P}_{2}+\bar{\bf A}_{2}+{\bf A}_{2}^{ex})
\phi_{\bf q}(2)\phi_{\bf p}(1)\;\; .
\end{equation}
The variation of the term $<H_{b}>$
leads to the following expressions:
\begin{equation}
H_{HF}^{3}\phi_{\bf p}(1)=-\frac{e}{mc}\sum_{\bf q}\int d2 \phi_{\bf
q}^{+}(2)
{\bf A}_{12}({\bf P}_{1}+\bar{\bf A}_{1}+{\bf A}_{1}^{ex})
\phi_{\bf q}(1)\phi_{\bf p}(2)\;\; ,
\end{equation}
\begin{equation}
H_{HF}^{4}\phi_{\bf p}(1)=-\frac{e}{mc}\sum_{\bf q}\int d2 \phi_{\bf
q}^{+}(2)
{\bf A}_{21}({\bf P}_{2}+\bar{\bf A}_{2}+{\bf A}_{2}^{ex})
\phi_{\bf q}(1)\phi_{\bf p}(2)\;\; ,
\end{equation}
\begin{equation}
H_{HF}^{5}\phi_{\bf p}(1)=-\frac{e^{2}}{mc^{2}}\sum_{{\bf q},{\bf k}}
\int d2 d3 \phi_{\bf
q}^{+}(2)\phi_{\bf k}^{+}(3)
{\bf A}_{32}{\bf A}_{31}\phi_{\bf q}(3)\phi_{\bf k}(2)\phi_{\bf p}(1)
\;\; .
\end{equation}
 Next terms:
\begin{equation}
H_{HF}^{6}\phi_{\bf p}(1)=\frac{e^{2}}{mc^{2}}\sum_{\bf q} \int d2
|\phi_{\bf q}(2)|^{2}|{\bf A}_{12}|^{2}\phi_{\bf p}(1)\;\; ,
\end{equation}
\begin{equation}
H_{HF}^{7}\phi_{\bf p}(1)=-\frac{e^{2}}{mc^{2}}\sum_{\bf q} \int d2
\phi_{\bf q}^{+}(2)|{\bf A}_{12}|^{2}\phi_{\bf q}(1)\phi_{\bf p}(2)\;\; ,
\end{equation}
are obtained from  the variations of $<H_{c}>$  and $<H_{d}>$,  respectively.
{}From the  variation of $<H_{e}>$
one has:
\begin{equation}
H_{HF}^{8}\phi_{\bf p}(1)=\frac{e^{2}}{mc^{2}}\sum_{{\bf q},{\bf k}}
\int d2 d3 \phi_{\bf
q}^{+}(3)\phi_{\bf k}^{+}(2)
{\bf A}_{12}{\bf A}_{13}\phi_{\bf q}(1)\phi_{\bf k}(3)\phi_{\bf p}(2)
\;\; ,
\end{equation}
\begin{equation}
H_{HF}^{9}\phi_{\bf p}(1)=\frac{e^{2}}{mc^{2}}
\sum_{{\bf q},{\bf k}}\int d2 d3 \phi_{\bf
q}^{+}(3)\phi_{\bf k}^{+}(2)
{\bf A}_{21}{\bf A}_{23}\phi_{\bf q}(1)\phi_{\bf k}(3)\phi_{\bf p}(2)
\;\; ,
\end{equation}
\begin{equation}
H_{HF}^{10}\phi_{\bf p}(1)=\frac{e^{2}}{mc^{2}}
\sum_{{\bf q},{\bf k}}\int d2 d3 \phi_{\bf
q}^{+}(3)\phi_{\bf k}^{+}(2)
{\bf A}_{31}{\bf A}_{32}\phi_{\bf q}(1)\phi_{\bf k}(3)\phi_{\bf p}(2)
\;\; .
\end{equation}
And the variation of the last  term leads to the expressions:
\begin{equation}
H_{HF}^{11}\phi_{\bf p}(1)=-\frac{e^{2}}{2mc^{2}}
\sum_{{\bf q},{\bf k}}\int d2 d3 \phi_{\bf
q}^{+}(2)\phi_{\bf k}^{+}(3)
{\bf A}_{12}{\bf A}_{13}\phi_{\bf q}(3)\phi_{\bf k}(2)\phi_{\bf p}(1)
\;\; ,
\end{equation}
\begin{equation}
H_{HF}^{12}\phi_{\bf p}(1)=-\frac{e^{2}}{mc^{2}}
\sum_{{\bf q},{\bf k}}\int d2 d3 \phi_{\bf
q}^{+}(2)|\phi_{\bf k}(3)|^{2}
{\bf A}_{31}{\bf A}_{32}\phi_{\bf q}(1)\phi_{\bf p}(2)
\;\; .
\end{equation}

In the following we will calculate  eigenvalues of
the single-particle Hamiltonian \\$H_{HF}=\sum_{i=1}^{12}H_{HF}^{i}$.
Let us consider first constant divergent terms:
\begin{equation}
\xi_{HF}^{6}=\frac{4f^{2}\hbar^{2}}{m}\rho \int d2 \frac{1}{|{\bf
r_{1}}-{\bf r}_{2}|^{2}} \; ,
\end{equation}
\begin{equation}
\xi_{HF}^{11}=-\frac{2f^{2}\hbar^{2}}{m}\frac{1}{\rho_{p}S}
\sum_{{\bf q}, {\bf k}}	\frac{1}{|{\bf k}-{\bf q}|^{2}}
\end{equation}
where $\rho_{p}=\frac{S}{(2\pi\hbar)^{2}}$.
Identifying $\pi\epsilon^{2}$ and $\pi\delta^{2}$ with the area per one
particle in the real  and the momentum space, respectively, we can
write ($\pi R^{2}=S$):
\begin{equation}
\xi_{HF}^{6}=4f^{2}E_{F}\lim_{R\rightarrow\infty}{\ln{\frac{R}{\epsilon}}}=
2f^{2}E_{F}\lim_{N\rightarrow\infty}{\ln{N}}\; ,
\end{equation}
\begin{equation}
\xi_{HF}^{11}=-2f^{2}E_{F}(\lim_{\delta\rightarrow 0}
{\ln{\frac{p_{F}}{\delta}}} -\frac{1}{2})=
-f^{2}E_{F}(\lim_{N\rightarrow\infty}\ln{N}-1)\;
\end{equation}
where $p_{F}$ is the Fermi momentum.
In $H_{HF}^{12}$ we recognize the logarithmic interaction exchange
term:
\begin{equation}
\xi_{HF}^{12}=- \frac{4f^{2}\hbar^{2}}{m}\frac{\rho}{\rho_{p}}
\sum_{{\bf q}\neq {\bf p}}	\frac{1}{|{\bf p}-{\bf q}|^{2}}\; .
\end{equation}
We have also:
\begin{equation}
\xi_{HF}^{1}=\frac{{\bf p}^{2}}{2m}\; ,
\end{equation}
\begin{equation}
\xi_{HF}^{2}=\xi_{HF}^{3}=\xi_{HF}^{4}=\xi_{HF}^{5}=0 \; .
\end{equation}
In remaining terms it is convenient to separate the cases of
 $|{\bf p}|<p_{F}$
 and $|{\bf p}|>p_{F}$. We obtain
\begin{equation}
\xi_{HF}^{7}(|{\bf p}|<p_{F})=2f^{2}\frac{{\bf
p}^{2}}{2m}-2f^{2}E_{F}\; ,\;\;
\xi_{HF}^{7}(|{\bf p}|>p_{F})=
4f^{2}E_{F}\ln{\frac{|{\bf p}|}{p_{F}}} \; ,
\end{equation}
\begin{equation}
\xi_{HF}^{8}(|{\bf p}|<p_{F})=\xi_{HF}^{9}(|{\bf p}|<p_{F})=
-2f^{2}\frac{{\bf p}^{2}}{2m}+ 2f^{2}E_{F}\; ,
\end{equation}
\begin{equation}
\xi_{HF}^{8}(|{\bf p}|>p_{F})=\xi_{HF}^{9}(|{\bf p}|>p_{F})= 0 \; ,
\end{equation}
\begin{equation}
\xi_{HF}^{10}(|{\bf p}|<p_{F})=2f^{2}\frac{{\bf p}^{2}}{2m}\; ,\;\;
\xi_{HF}^{10}(|{\bf p}|>p_{F})=2f^{2}E_{F}\frac{p_{F}^{2}}{{\bf p}^{2}}\; .
\end{equation}
Finally we can write:
\begin{equation}
\label{e1}
\xi_{HF}(|{\bf p}|<p_{F})=\frac{{\bf p}^{2}}{2m}+3f^{2}E_{F} +
f^{2}E_{F}\lim_{N\rightarrow\infty}{\ln{N}}
- \frac{2f^{2}E_{F}}{\pi\rho_{p}}
\sum_{{\bf q}\neq {\bf p}}	\frac{1}{|{\bf p}-{\bf q}|^{2}}\;
\end{equation}
and
\begin{equation}
\label{e2}
\xi_{HF}(|{\bf p}|>p_{F})=\frac{{\bf p}^{2}}{2m}+2f^{2}E_{F}
(2\ln{\frac{|{\bf p}|}{p_{F}}}+\frac{1}{2}+\frac{p_{F}^{2}}{{\bf p}^{2}})
+f^{2}E_{F}\lim_{N\rightarrow\infty}{\ln{N}}
- \frac{2f^{2}E_{F}}{\pi\rho_{p}}
\sum_{{\bf q}\neq {\bf p}}	\frac{1}{|{\bf p}-{\bf q}|^{2}}\; .
\end{equation}
Let us now perform explicitly the sum in Eq. (\ref{e1}). One finds:
\begin{equation}
\sum_{{\bf q}\neq {\bf p}}	\frac{1}{|{\bf p}-{\bf
q}|^{2}}=\pi\rho_{p}\lim_{\delta\rightarrow 0}
(\ln{\frac{p_{F}-p}{\delta}}+\ln{\frac{p_{F}+p}{\delta}})
\end{equation}
For $|{\bf p}|>p_{F}$ the sum equals:
\begin{equation}
\label{integral}
\sum_{{\bf q}\neq {\bf p}}\frac{1}{|{\bf p}-{\bf
q}|^{2}}=4\rho_{p}\int_{0}^{\alpha}\psi
\frac{p\sin{\psi}}{\sqrt{p_{F}^{2}-p^{2}(\sin{\psi})^{2}}} d\psi
\end{equation}
where $\sin{\alpha}=\frac{p_{F}}{p}$.
To see the most important feature of the single-particle energy spectrum
it is enough to use the following estimation:
\begin{equation}
\int_{0}^{\alpha}\psi
\frac{p\sin{\psi}}{\sqrt{p_{F}^{2}-p^{2}(\sin{\psi})^{2}}} d\psi
<\int_{0}^{\alpha}\frac{\pi}{2}
\frac{p\sin{\psi}}{\sqrt{p_{F}^{2}-p^{2}(\sin{\psi})^{2}}} d\psi
=\frac{\pi}{4}\ln{\frac{p+p_{F}}{p-p_{F}}}.
\end{equation}
If only $(p_{F}-p')>0$ and $(p''-p_{F})>0$  are
 not infinitesimal  the dominating contribution to the energy
difference $\xi(p'')-\xi(p')$ is greater than
\begin{equation}
\label{ediff}
\xi(p'')-\xi(p')>	2f^{2}E_{F}\lim_{\delta\rightarrow 0}
(\ln{\frac{p_{F}+p'}{\delta}}+
\ln{\frac{p_{F}-p'}{\delta}})
\end{equation}
which is divergent. Hence, there is an infinite	jump of the single-particle
energy at the Fermi momentum.

An infinite gap (or jump) in the single-particle energy
spectrum  appears to be the characteristic
feature of the Hartree-Fock ground state of Chern-Simons systems
\cite{HFA,fqhe}.
However, in the present case this feature has no physical meaning due
to the screening of the logarithmic interaction in the RPA \cite{HLR}.
This makes the physics distinct from that of the anyon superconductor
where the logarithmic interaction remains unscreened and
single-particle excitations are vortices \cite{Dai}.

In
the composite fermion metal  the RPA Fermi velocity vanishes \cite{HLR}.
One can see in Eqs. (\ref{e1}, \ref{e2}) that
the only H-F contributions to the Fermi velocity comes from $\xi_{HF}^{1}$
and $\xi_{HF}^{12}$. However, the free term $\xi_{HF}^{1}$
and the logarithmic interaction exchange term
$\xi_{HF}^{12}$ were included in the analysis of HLR, thus the
Hartree-Fock result brings no corrections to the HLR conclusions.

\vspace{0.5cm}
\noindent{\bf 3. Conclusions}
\vspace{0.5cm}

The Hartree-Fock single-particle energy spectrum of the
composite fermion metal
is found. The dominating contribution is the
logarithmic interaction exchange term which produces an infinite
jump of the single-particle energy at the Fermi momentum.
However, this feature has no
physical meaning due to the
screening of the logarithmic interaction in the RPA.
We find also that the Hartree-Fock terms not included in
the RPA bring no
corrections to the Fermi velocity.

\end{document}